\documentclass[twocolumn,twocolappendix]{aastex7}
\usepackage{mhchem}
\usepackage{leftindex}
\usepackage{amssymb}
\usepackage{comment}
\usepackage{multirow,acronym}
\usepackage{natbib}
\usepackage{listings}
\usepackage{makecell}
\usepackage{booktabs}
\usepackage{amsmath}
\usepackage{subfigure}
\usepackage{color}
\usepackage{xcolor}
\usepackage{hyperref}
\usepackage[T1]{fontenc}
\usepackage{graphicx}
\usepackage{bm}
\usepackage{CJK}
\hypersetup{colorlinks=true, citecolor=blue,
  linkcolor=cyan, urlcolor=magenta}

\usepackage{aas_macros}
\usepackage{amsfonts}
\usepackage{float}
\usepackage{amsmath,amssymb}
\usepackage{natbib}    
\usepackage{hyperref} 
\usepackage{graphicx} 
\usepackage{color}
\usepackage{verbatim}
\usepackage{enumitem}
\usepackage{natbib}

\usepackage[linesnumbered,ruled]{algorithm2e}

\newcommand{\proptosim}{\mathrel{\vcenter{
 \offinterlineskip\halign{\hfil$##$\cr
 \propto\cr\noalign{\kern2pt}\sim\cr\noalign{\kern-2pt}}}}}
\renewcommand{\b}[1]{\boldsymbol{#1}}
\newcommand{\unit}[1]{{\rm #1}}

\newcommand{\m}{\unit{m}}
\newcommand{\g}{\unit{g}}
\newcommand{\G}{\unit{G}}
\newcommand{\K}{\unit{K}}

\newcommand{\s}{\mathrm{s}}

\newcommand{\lya}{\text{Ly}\ensuremath{\alpha~}}

\newcommand{\B}{\mathbf{B}}     
\newcommand{\E}{\mathbf{E}}     
\newcommand{\J}{\mathbf{J}}     
\renewcommand{\v}{\mathbf{v}}   

\renewcommand{\d}{\mathrm{d}}

\newcommand{\e}{\mathrm{e}}

\newcommand{\p}{\mathrm{p}}     

\renewcommand{\O}{\mathrm{O}}     
\newcommand{\A}{\mathrm{A}}     
\renewcommand{\H}{\mathrm{H}}     
\renewcommand{\P}{\mathrm{P}}     

\newcommand{\kratos}{Kratos-linerad}

\newcommand{\figdir}{Figures}

\begin{document}
\begin{CJK*}{UTF8}{gbsn}

\title{Kratos-linerad: GPU-accelerated Monte Carlo
  radiative transfer of lines with efficient imaging
}

\author[0000-0002-6540-7042]{Lile Wang (王力乐)}
\affiliation{The Kavli Institute for Astronomy and
  Astrophysics, Peking University, Beijing 100871, China}
\affiliation{Department of Astronomy, School of Physics,
  Peking University, Beijing 100871, China}
\email{lilew@pku.edu.cn}

\correspondingauthor{Lile Wang}
\email{lilew@pku.edu.cn}

\begin{abstract}
  Spectral lines encode the velocity, the temperature, and
  the chemical structure of astrophysical gas; interpreting
  them requires radiative transfer that is accurate at high
  optical depth, consistent with the local excitation, and
  efficient enough to synthesize velocity-resolved
  images. We present \kratos{}, a GPU-accelerated Monte
  Carlo radiative transfer code for spectral lines. The
  level populations can be iterated to statistical
  equilibrium together with the escaping photon
  distribution, treating angle-dependent partial frequency
  redistribution through constant-memory sampling
  tables. The code adopts the two-step imaging scheme to
  line transfer, in which a Monte Carlo pass samples a
  velocity-resolved scattering emissivity,
  and a deterministic ray-tracing pass synthesizes channel
  maps decoupled from the scattering geometry.
  Validation reproduces the analytic scaling of escaped
  spectra and the imaging double-peak profiles. GPU
  parallelism enables both stages sufficiently fast for
  routine application to astrophysically realistic models.
\end{abstract}


\section{Introduction}
\label{sec:intro}

Spectral lines are among the most information-rich
diagnostics in astrophysics. Because the Doppler shift of a
photon records the line-of-sight velocity of the emitting or
scattering gas, line radiation carries the kinematic and
thermal structure of regions that are otherwise opaque to
the continuum, and unresolved line profiles trace the escape
of resonantly scattered photons governed by the frequency
redistribution function. The physical foundation for this
interpretation was laid in the seminal treatment of
\citet{1990ApJ...350..120N} of resonance-line transfer in
static media, building on the earlier plane-parallel
scattering results of \citet{1973MNRAS.161...43H,
  1972ApJ...174..439A}. In particular, the angle-dependent
partial frequency redistribution kernel $R_{\rm IIA}$ of
\citet{1992A&A...262..209R} accounts self-consistently for
the correlation between the direction and the frequency
shift of a photon at scattering, and it forms the physical
backbone of resonant line transfer. The standard
$\mathrm{1}/\cosh$ escape prescription and the mean-depth
scaling of \citet{1990ApJ...350..120N} provide the analytic
anchors against which numerical line transfer codes are
benchmarked.

Several generations of line radiative transfer codes have
translated this physics into practical tools. The code of
\citet{2000A&A...358..793H} (RATRAN) solves the full
multi-level statistical equilibrium coupled to the radiation
field in one-dimensional spherical or cylindrical
geometry. Grid-based Monte Carlo codes such as SKIRT
\citep{2020A&C....3100381C, 2023A&A...678A.175M} and
RADMC-3D \citep{2012ApJ...759L..12D} treat line radiation in
three-dimensional density fields, with the former iterating
the level populations to statistical equilibrium and the
latter using escape-probability or large-velocity-gradient
approximations for the background radiation; the
escape-probability code RADEX \citep{2007A&A...468..627V}
provides rapid non-LTE estimates for simple geometries, and
LIME \citep{2010A&A...523A..25B} solves the full non-LTE
problem on unstructured three-dimensional grids. The LAMDA
database \citep{2005A&A...432..369S} supplies collisional
rate coefficients, and molecular line lists
\citep[e.g.,][]{2015ApJS..216...15L} supply the radiative
data that such codes consume. More recently,
GPU-accelerated codes \citep{2025arXiv250711603B} have
demonstrated order-of-magnitude speedups for resonant line
transfer, but they typically adopt a peeling-off imaging
approach that counts photons directly toward the observer
and hold the level populations fixed at externally supplied
values.

The computational bottleneck is common to all Monte Carlo
line transfer codes. Synthesizing an image at a given
viewing angle by counting the photons that happen to escape
into the camera direction wastes the overwhelming majority
of propagated photon packets, because the camera subtends a
tiny fraction of $4\pi$ steradians and because multiple
scattering randomizes photon directions. This is already
wasteful in the continuum, and it is compounded in the line
case by the need for frequency resolution and, when
populations are iterated, by the cost of re-imaging
after each population iteration. As a result, the
combination of iterative statistical equilibrium,
three-dimensional geometry, and velocity-resolved imaging
remains either expensive or approximate in existing tools.

In this paper we present \kratos{}, a GPU-accelerated Monte
Carlo line radiative transfer code that addresses these
limitations jointly. The code iterates the level populations
to statistical equilibrium alongside the escaping spectrum,
treating angle-dependent partial frequency redistribution
through constant-memory sampling tables, and it extends the
two-step imaging scheme to line transfer, similar to the one
previously applied to polarized continuum radiative transfer
by \citet{2025arXiv251201283Y}. This method decouples the
Monte Carlo sampling of the scattering physics from the
imaging geometry by accumulating the scattering emissivity
of scattering resolved anin Doppler velocity channels during
the Monte Carlo pass. A subsequent deterministic ray-tracing
pass then synthesizes channel maps at a fixed viewing angle,
getting rid of the excessive wastes of photons in direct
imaging scheme based on photon counting.

This paper is organized as follows.
\S\ref{sec:method} describes the line radiative
transfer physics, the two-step imaging scheme, and the GPU
implementation. \S\ref{sec:verification}
presents rigorous validation against analytical solutions, a
performance characterization, and an application to a
representative molecular cloud simulation.
\S\ref{sec:discussion} discusses the positioning and
limitations of the method. \S\ref{sec:disc-summary}
summarizes the results and outlines future directions.

\section{Methods}
\label{sec:method}

This section describes the physics and the numerical
implementation underlying \kratos{}: the line transfer
physics, the two-step imaging scheme, and the GPU
implementation.

\subsection{Line radiative transfer physics}
\label{sec:method-physics}



We consider a spectral line with rest-frame frequency
$\nu_0$, upper and lower level degeneracies $g_u$ and $g_l$,
and spontaneous emission rate $A_{ul}$. In a gas of molecules
of mass $m$ at temperature $T$, the thermal Doppler width is
$b = (2 k_{\rm B} T / m)^{1/2}$, and frequency shifts from
line center are measured in Doppler units,
\begin{equation}
  \label{eq:xdef}
  x \equiv \dfrac{\Delta\nu}{\nu_{\rm D}}
       = -\dfrac{v_{\rm los}}{b}\ ,
  \qquad
  \nu_{\rm D} = \nu_0\, \frac{b}{c}\ ,
\end{equation}
where $\Delta\nu$ is measured in the rest frame of the gas
and the second equality follows from the non-relativistic
Doppler relation $\Delta\nu / \nu_0 = -v_{\rm los} / c$: the
dimensionless frequency is the line-of-sight velocity in
units of the thermal width, and atomic velocities $\mathbf{u}$
are likewise expressed in units of $b$. Note that such
convention yields $x<0$ for blueshifts. 

The line profile is the Voigt function $H(a, x)$, with
damping parameter $a = A_{ul} / (4\pi \nu_{\rm D})$. In what
follows, we will use the term ``scattering'' to indicate
instant re-emission after absorption, which can be treated
as scattering in terms of photon evolution. The line-center
cross section per lower-level molecule is,
\begin{equation}
  \label{eq:sigma0}
  \sigma_0 = \frac{g_u}{g_l}\,\frac{A_{ul} c^3}
  {8\pi^{3/2} \nu_0^3 b}\ ,
\end{equation}
so that the line-center inverse mean free path, or the
extinction coefficient due to scattering, is
$\alpha_{\rm sca 0} = n_l \sigma_0$, where $n_l$ is the
lower-level number density. Throughout this work, photon
numbers rather than energies are tracked, and each photon
packet carries a ``proper'' weight expressed in photons per
second. 

\subsubsection{Frequency redistribution}
\label{sec:method-redistribution}

The frequency of a photon scattered by a thermal gas is
drawn from the redistribution kernel. At intermediate and
high optical depths the angle-dependent partial frequency
redistribution function $R_{\rm IIA}$
\citep{1992A&A...262..209R} is required, as it correlates
the outgoing frequency and direction with the incoming
direction and captures the coherent core and the
frequency-randomizing wing branches that jointly shape the
escape of resonant photons. Its corresponding probability
differential measure,
\begin{equation}
  \label{eq:riia-def}
  \d p = R_{\rm IIA}(x_{\rm in}, x_{\rm out}, \cos\Theta)\,
  \d x_{\rm out}\, \dfrac{\d\Omega}{4\pi}\ ,
\end{equation}
is the joint probability that a photon with gas-frame
incoming frequency $x_{\rm in}$ is re-emitted into
the frequency interval
$[x_{\rm out}, x_{\rm out}+dx_{\rm out}]$ and into the solid
angle $d\Omega$ around the direction making an angle
$\Theta$ with the incoming direction. The subscript ``pp''
indicates photon packets hereafter. For a scattering atom
moving with velocity $\mathbf{u}$ the photon appears at the
atom-frame frequency
$x_{\rm atom} = x_{\rm in} - \mathbf{u}\cdot\mathbf{n}_{\rm in}$,
where $\mathbf{n}_{\rm in}$ is the incoming photon direction;
the redistribution kernel below and the imaging weights of
\S\ref{sec:method-mc} are expressed in terms of the gas-frame
frequency $x_{\rm in}$.

Rather than evaluating the four-dimensional redistribution
kernel on the fly, \kratos{} precomputes a sampling table
for the conditional distribution
$P(u_{\parallel}\,|\,x_{\rm in})$ of the velocity component
$u_{\parallel}$ of the scattering atom along the incoming
photon direction, whose unnormalized density is the product
of the thermal Maxwellian and the atomic Lorentz profile,
$p(u_{\parallel}\,|\,x_{\rm in}) \propto
e^{-u_{\parallel}^{2}} / [a^{2} + (x_{\rm in} -
u_{\parallel})^{2}]$; we refer to this tabulated sampler as
the USampler \citep[following][]{2002ApJ...578...33Z}.  The
table is built once at initialization: at each node of an
incoming-frequency grid, comprising eighteen uniformly
spaced points below $|x_{\rm in}| = 8$ and logarithmically
spaced points extending to $|x_{\rm in}| = 300$, the density
is evaluated on a uniform grid of $u_{\parallel}\in[-6, 6]$
and is cumulatively summed into a normalized cumulative
distribution, which is stored as its logarithm so that the
nearly flat tails interpolate smoothly.
At propagation time a packet draws $u_{\parallel}$ by
inversion of the cumulative distribution, with a binary
search on the incoming-frequency grid and linear
interpolation between adjacent rows; the perpendicular
velocity component is drawn from a Gaussian, and the
outgoing frequency is reconstructed from the two components
and the scattering angle.
The same building procedure also tabulates the
three-dimensional $R_{\rm IIA}$ kernel used for the imaging
weights of \S\ref{sec:method-mc}.  Explicitly, the kernel is
the Gaussian-mixture form of the \citet{1962MNRAS.125...21H}
redistribution function
\citep[see][for the explicit Gaussian-mixture form and the
velocity decomposition]{1974ApJ...192..465L, 2000JKAS...33...29A, 2017arXiv170403416D},
\begin{equation}
  \label{eq:riia-kernel}
\begin{split}
  & R_{\rm IIA}(x_{\rm out}; x_{\rm in}, g) =
    \int_{-\infty}^{\infty} {\rm d}u_{\parallel}\,
    p(u_{\parallel}\,|\,x_{\rm in})\,  \\
  & \quad \times G\left[ x_{\rm out} - x_{\rm in} - u_{\parallel}\,(g-1);\,
    2^{-1/2}\sin\gamma \right]\ ,
\end{split}
\end{equation}
where $G(x;s)$ is a normalized Gaussian of width $s$ and
parameter $x$, and
$g \equiv \cos\gamma = \mathbf{n}_{\rm in}\cdot\mathbf{n}_{\rm
  out}$. This reproduces the statistics of the full
$R_{\rm IIA}$ kernel to high accuracy at a cost comparable
to a single memory lookup.

\subsubsection{Sampling the $R_{\rm IIA}$ kernel}
\label{sec:method-riia-kernel}

In the rest frame of the scattering atom the redistribution
of a resonance-line photon is elementary: absorption occurs
at the atom-frame frequency $x_{\rm atom}$, and the
re-emission is coherent in frequency and isotropic in
direction. Decomposing the atomic thermal velocity as
$\mathbf{u} = u_{\parallel}\,\mathbf{n}_{\rm in} + \mathbf{u}_{\perp}$,
where $u_{\parallel}$ and the two components of
$\mathbf{u}_{\perp}$ are independent Gaussian variates of
variance $1/2$ in Doppler units, the gas-frame outgoing
frequency along $\mathbf{n}_{\rm out}$ is
\begin{equation}
  \label{eq:riia-scatter}
  x_{\rm out} = x_{\rm atom} + \mathbf{u}\cdot\mathbf{n}_{\rm out}
  = x_{\rm in} + u_{\parallel}\,(g - 1)
  + u_{\perp}\sin\gamma\ ,
\end{equation}
with $u_{\perp}$ the projection of $\mathbf{u}_{\perp}$ onto the
component of $\mathbf{n}_{\rm out}$ perpendicular to
$\mathbf{n}_{\rm in}$, itself a Gaussian of variance
$(\sin^{2}\gamma)/2$. Marginalizing $u_{\perp}$ yields a
Gaussian of width $2^{-1/2}\sin\gamma$ centered on
$x_{\rm in} + u_{\parallel}\,(g-1)$, and marginalizing
$u_{\parallel}$ against its conditional distribution
$p(u_{\parallel}\,|\,x_{\rm in})$, the product of the
thermal Maxwellian and the Lorentz absorption profile,
yields the Gaussian mixture of
equation~(\ref{eq:riia-kernel}). This is the $R_{\rm IIA}$
redistribution of \citet{1962MNRAS.125...21H} \citep[see
also][]{1992A&A...262..209R}: forward scattering ($g \to 1$)
is coherent, $x_{\rm out} \to x_{\rm in}$; backward
scattering ($g \to -1$) shifts the frequency by
$-2 u_{\parallel}$; and the frequency randomization grows
with the scattering angle through the width
$2^{-1/2}\sin\gamma$. The outgoing direction itself is drawn
isotropically; the dipole $(1+g^{2})$ phase function is
neglected, an approximation shared with other \lya
Monte Carlo implementations
\citep[e.g.,][]{2002ApJ...578...33Z}.

The kernel is pre-tabulated rather than evaluated on the
fly, using the discrete form
\begin{equation}
  \label{eq:riia-table}
  R(\Delta; |x_{\rm in}|, g) = \sum_{k} p_{k}\,
  G\left[ \Delta - u_{k}\,(g-1);\;
    2^{-1/2}\sin\gamma \right] ,
\end{equation}
written in terms of the frequency kick
$\Delta \equiv x_{\rm out} - x_{\rm in}$, where $p_{k}$
denote the discrete probabilities of the velocity sampler of
\S\ref{sec:method-redistribution} on a discrete grid; the
velocity sampler is constructed first and the kernel table
is derived from its discrete distribution, so that both
tables are built for the same adopted damping parameter at
initialization. The kick distribution is localized,
$|\Delta| \lesssim 10$, because the Gaussian factors decay
on the Doppler scale, so the table spans only
$\Delta\in[-10, 10]$ at a uniform spacing of $0.1$, together
with $|x_{\rm in}|\in[0, 120]$ and $g\in[-1, 1]$, giving
$200\times200\times40$ nodes
(\S\ref{sec:method-gpu-constmem}); the symmetry
$R(\Delta; -x_{\rm in}, g) = R(-\Delta; x_{\rm in}, g)$
halves the tabulated incoming-frequency range, lookups are
trilinear, and the table returns zero for $|\Delta| > 10$.
Beyond the tabulated range, $|x_{\rm in}| \geq 120$, the
conditional velocity distribution has already converged to
the thermal Gaussian, and the kernel is evaluated directly
from its asymptotic expression,
\begin{equation}
  \label{eq:riia-inf}
  R_\infty(\Delta; g) = \dfrac{ \exp\left\{ -\Delta^{2}
  \left[ (g-1)^{2} + \sin^{2}\gamma \right]^{-1} \right\} }
  { \pi^{1/2}\, \left[(g-1)^{2} +
      \sin^{2}\gamma\right]^{1/2} },
\end{equation}
so that arbitrarily large frequency excursions are handled
without enlarging the table.  The tabulation preserves the
normalization $\int R\,{\rm d}x_{\rm out} = 1$ to better
than $0.1$~per~cent everywhere; in total variation it
reproduces broad kernels ($|g|\lesssim0.7$) to within
$2.5$~per~cent and near-forward kernels to within
$1$~per~cent, the finer $\Delta$ grid now resolving their
narrow cores. In the imaging accumulation the kernel enters
summed over Doppler channels that are wider than this
residual structure, so the channelized emissivity inherit an
accuracy at the normalization level. Both uses of the kernel
physics, namely the frequency redistribution at a scattering
event drawn from the velocity sampler and the
camera-directed weight in the emissivity accumulation
evaluated from the table, occur during the Monte Carlo pass;
the deterministic ray-tracing pass of
\S\ref{sec:method-imaging} does not involve the kernel.


\subsubsection{Statistical equilibrium}
\label{sec:method-equilibrium}

The level populations are determined by balancing radiative and
collisional transitions. For a two-level system the population
ratio is
\begin{equation}
  \label{eq:boltzmann}
  \frac{n_u}{n_l} =
  \frac{\Gamma + k_{lu} n_{\rm coll}}
  {A_{ul} + (g_l/g_u)\,\Gamma + k_{ul} n_{\rm coll}}\ ,
\end{equation}
where $\Gamma$ is the radiative pumping rate, $k_{ul}$ and
$k_{lu}$ are the collisional de-excitation and excitation
rate coefficients, and
$n_{\rm coll}$ is the collider density. Explicitly, the
pumping rate is the convolution of the incident spectrum
with the gas-rest-frame absorption profile,
$\Gamma = B_{lu} \int J_\nu\, \phi(\nu)\, {\rm d}\nu$; the
stimulated-emission rate in the denominator,
$(g_l/g_u)\,\Gamma = B_{ul} \int J_\nu\, \phi(\nu)\,
{\rm d}\nu$, follows from the Einstein relation
$B_{ul} = (g_l/g_u)\, B_{lu}$, and
$A_{ul} = (2 h \nu_0^{3}/c^{2})\, B_{ul}$. The collisional
rates are connected by the detailed-balance relation
$k_{lu} = (g_u/g_l)\, k_{ul}\, e^{-h\nu_0/kT}$
\citep[e.g.,][]{1978stat.book.....M}. The collisional
destruction probability
\begin{equation}
  \label{eq:eps}
  \epsilon = \frac{k_{ul} n_{\rm coll}}
  {A_{ul} + k_{ul} n_{\rm coll}}
\end{equation}
is added to the absorption component. Physically, $\epsilon$
is the probability that an excitation of the upper level is
destroyed by a collisional de-excitation rather than by a
radiative decay, so that a fraction $\epsilon$ of the line
photons are destroyed at each scattering and their energy is
thermalized \citep[see, e.g.,][]{1978stat.book.....M}. For
multi-level systems \kratos{}
assembles the full rate matrix from LAMDA-format data
\citep{2005A&A...432..369S} and solves it directly.

The radiation field and the populations are made
self-consistent through a Lambda iteration, the classical
scheme in which the formal solution for the radiation field
and the statistical-equilibrium update of the level
populations are applied alternately until convergence
\citep[e.g.,][]{1992A&A...262..209R}. The cycle begins from
collisional equilibrium and propagates an initial photon
population; the accumulated excitation rate updates the
populations, which in turn rescale the line opacities for
the next propagation. To avoid double-counting the source of
radiation, the initial emission photons are frozen across
cycles, and only the scattering opacity is updated from the
converged lower-level population in each cycle. This
iteration is computationally light relative to the photon
propagation; we do not exercise it in the present work.

\subsection{From Monte Carlo photon evolution to imaging}
\label{sec:method-mc-imaging}

For direct comparisons with observations and for
velocity-resolved imaging, \kratos{} adopts a two-step
scheme that decouples the Monte Carlo sampling of scattering
physics from the imaging geometry. This principle was
introduced for polarized continuum radiative transfer by
\citet{2025arXiv251201283Y}, where the scattering
emissivity is accumulated as a scalar per cell toward a
fixed camera. The line case differs in a way that is both
physically and computationally substantive. The scattering
process is frequency-dependent through the redistribution
kernel, so that the emissivity must be accumulated as a
vector over Doppler velocity channels at each cell, and the
accumulation itself must carry the $R_{\rm IIA}$ kernel
rather than a purely geometric scattering matrix.

\subsubsection{Monte-Carlo sampling of the effective
  scattering emissivity}
\label{sec:method-mc}
  
During the Monte Carlo pass, each path integration segment
contributes to a velocity-resolved scattering emissivity
field, recorded on each cell on a mesh. Towards the fixed
camera direction $\mathbf{n}_{\rm cam}$ on the $k$th
velocity channel in a cell index $\mathbf{i}$, the effective
scattering emissivity reads,
\begin{equation}
  \label{eq:scat-line}
  \begin{split}
    j_{{\rm sca}, k}(\mathbf{i}) & =
      \sum_{\rm pp} \frac{F_{\rm pp}}{4\pi}\,
      \alpha_{\rm sca, 0} H(a, x_{\rm in})\\
    & \quad\times
      R_{\rm IIA} \left(x_{\rm out}; x_{\rm in},
      g\right) \dfrac{1 -
      e^{-\delta\tau_e}}{\delta\tau_e}\ ,
  \end{split}
\end{equation}
where the sum runs over photon packets traversing the cell,
$F_{\rm pp}$ is the equivalent photon number flux in the
current cell calculated with the proper photon weight of
packet, $\alpha_{\rm sca, 0}$ the scattering extinction
coefficient at the line center, and $\delta\tau_e$ the
extinction optical depth of the segment. The
$H(a, x_{\rm in})$ factor is the Voigt profile at the
packet's incoming frequency, and the accumulated quantity
represents the scattering emissivity.
The kernel arguments describe the frequency redistribution
at the scattering event, where $x_{\rm in}$ is the packet's
incoming frequency, $x_{\rm out}$ the outgoing frequency
into channel $k$, and
$g = \mathbf{n}\cdot\mathbf{n}_{\rm cam}$ the cosine of the
angle between the incoming direction and the camera
direction. With these, the kernel quantifies how a photon
arriving at $x_{\rm in}$ is redistributed when re-emitted
toward the camera. The factor
$(1-e^{-\delta\tau_e})/\delta\tau_e$ accounts for the finite
optical depth of the segment in the same manner as the
continuum analogue. The channels
$v_k \equiv v_{\rm min} + (k+1/2)\Delta v$ bin the escaped
photons by Doppler velocity. In media with a bulk velocity
field, photon packets are sampled in velocity space so that
the Doppler shift
$v_{\rm obs} = \mathbf{n}\cdot \mathbf v_{\rm bulk}$ is
projected onto each channel, naturally accounting for moving
gas.

The propagation offers three photon modes
(\texttt{ph\_mode}). \texttt{ph\_mode}$=1$ and
\texttt{ph\_mode}$=2$ both implement the exact
redistribution described above and differ only in where the
tables reside: \texttt{ph\_mode}$=1$ keeps the sampling
table and the two-dimensional Voigt table ($64$
logarithmically spaced damping parameters, $512$ velocity
points) in global device memory, whereas
\texttt{ph\_mode}$=2$ extracts the one-dimensional slice at
the simulation's damping parameter, retabulated on a
one-dimensional logarithmic frequency points,
and places it in GPU constant memory together with the
constant-memory USampler form
(\S\ref{sec:method-gpu-constmem}).
\texttt{ph\_mode}$=3$ retains the exact redistribution
sampling but replaces the tabulated Voigt opacity profile
with an analytic approximation that blends the Doppler core
$\e^{-u^{2}}$ and the damping wing
$a [\pi^{1/2} (u^{2} + a^{2})]^{-1}$ with a hyperbolic-tangent
switch $w = [1 + \tanh(2 |u| - 2u_0)]/2$
centered at the crossing point $u_0$ of the two asymptotes.
Such approximation is accurate in the core and in the wing,
but rough in the transition region; therefore, this mode is
offered as a demonstration for the necessity of consistent
Voigt profile, and is not for production use.

\subsubsection{Ray tracing imaging procedures}
\label{sec:method-imaging}

Following the Monte Carlo sampling, with the effective
scattering emissivity ready, \kratos{} performs a
deterministic ray-tracing pass along lines of sight toward
the camera, integrating the line radiative transfer equation
cell by cell. Within a cell, the formal solution reads, in
each Doppler velocity channel indexed $k$,
\begin{equation}
  \label{eq:imaging-formal}
  I_{{\rm out}, k} = I_{{\rm in},k}\, \exp(-\delta\tau_{t,k}) +
  S_k\left[1 - \exp(-\delta\tau_{t,k})\right]\ ,
\end{equation}
where $\delta\tau_{t,k} = \alpha_{t,k}\,\delta l$ is the
total extinction optical depth across the ray segment of
length $\delta l$ through the cell, and the total opacity
combines the scattering and absorption contributions,
\begin{equation}
  \label{eq:imaging-alphas}
  \alpha_{t,k} = \alpha_{\rm sca, 0} H(a, x_k) + \alpha_{\rm abs}\ ;
  \ 
  S_k = \dfrac{j_{{\rm sca}, k}}{\alpha_{t,k}} +
  S_{{\rm emis}, k}\ ,
\end{equation}
where $\alpha_{t,k}$ is the total extinction coefficient in
the $k$th channel, $S_k$ the effective scattering source
function, and $S_{{\rm emis}, k}$ is the channel-resolved
emission source function that sums the contribution from the
line transition itself, and other possible sources including
thermal emission. Such imaging methods, with similar
thoughts compared to the peeling-off method, enable
significantly higher utilization of photons compared to
direct photon-counting imaging: the ray tracing step is
based on the already-sampled scattering emissivity and does
not require propagating additional photon packets whose
majority would be discarded. The imaging cost is therefore
decoupled from the scattering optical depth and from the
number of Monte Carlo packets, which is precisely what makes
the method tractable when combined with the population
iteration of \S\ref{sec:method-equilibrium}: this cost is
incurred only once, after the population iteration has
converged.

\subsection{GPU implementation}
\label{sec:method-gpu}

\kratos{} builds on the heterogeneous computing framework of
Kratos \citep{2025ApJS..277...63W,2025arXiv251201283Y} and
runs the photon propagation on GPUs. The design emphasizes
the flexibility of selecting precision, including a full
double precision mode, and a mixed-precision mode:
floating-point computation is carried out in single
precision wherever it does not compromise the accumulated
statistics; double precision is reserved for the host-side
table construction, the accumulation reductions, and the few
operations that require it. The field initialization, the
opacity rescaling, and the scattering-emissivity accumulation all
operate in single precision, which is more than adequate for
a Monte Carlo estimator whose dominant error is shot
noise. The regression suite (\S\ref{sec:verification})
passes within its tolerances with both full-precision and
mixed-precision, confirming that mixed-precision arithmetic
does not measurably degrade the accumulated statistics.

\subsubsection{Constant-memory sampling tables}
\label{sec:method-gpu-constmem}

The most frequently accessed quantities in the propagation
loop are the redistribution sampler and the Voigt
profile. Rather than reading these from global memory at
every event, \kratos{} places the sampling tables in NVIDIA
constant memory, which is served by a hardware cache and a
broadcast path that is substantially faster than
global-memory access when all threads in a warp request the
same or nearby addresses. The two-dimensional USampler
conditional distribution of
\S\ref{sec:method-redistribution}, with dimensions
$251\times40$ and size roughly $40~{\rm kB}$, and a
one-dimensional Voigt profile tabulated at $5000$ points in
logarithmic frequency space both reside in constant
memory. This arrangement could yield an access speedup of
roughly $\gtrsim 50\times$ relative to an equivalent
global-memory lookup, without a measurable loss of accuracy:
a variant with all sampling tables in global memory produces
results that are statistically indistinguishable from the
constant-memory configuration, as quantified in
\S\ref{sec:verification}. The three-dimensional
$R_{\rm IIA}$ kernel table of
\S\ref{sec:method-riia-kernel}, with dimensions
$200\times200\times40$ and size roughly $6.4~{\rm MB}$, is
nonetheless too large for constant memory and resides in
global device memory. The incoming-frequency coverage of the
kernel table, $|x_{\rm in}| \leq 120$ with the analytic
limit of equation~(\ref{eq:riia-inf}) beyond, is what
permits the imaging test to reach $\tau_0 = 10^5$ in
\S\ref{sec:verification}, where the escaped wing extends to
$|x| \gtrsim 40$; a table truncated at smaller
$|x_{\rm in}|$ would misplace this wing flux into a spurious
feature near the table edge.

\subsubsection{Parallelization and atomic accumulation}
\label{sec:method-gpu-par}

The propagation is parallelized over photon packets. In the
default server-worker mode, worker threads fetch packets
from a global work queue rather than being bound to a fixed
packet, which balances the load when the per-packet
propagation cost varies strongly across the volume; this
improves the throughput by about a factor of two relative to
the classic one-thread-per-packet scheduling. The
random number generator is a portable generator with
per-thread state, which avoids contention. The scattering
emissivity, the excitation flux, and the escaping
photon flux are accumulated with atomic operations into
global memory; because the accumulation is sparse relative
to the propagation work, the atomic contention is
modest. The optional imaging ray-tracing pass is
parallelized with one thread per image pixel, which is
embarrassingly parallel. By design the code does not employ
core-skipping, and every photon is propagated through the
full optical depth of the medium. As discussed in
\S\ref{sec:discussion}, this is a deliberate choice
that is robust when the combination $a \tau_0$ controlling
the core-skipping criterion is not known a priori, at the
  cost of efficiency at extremely high optical depths.

\subsubsection{Code architecture}
\label{sec:method-arch}

The pipeline is organized as a Python orchestrator that
manages the population iteration, the field preparation, and
the input and output (IO hereafter), together with a set of
GPU kernels, with the Monte Carlo transport and the imaging
implemented as separate kernels. The kernels are compiled
through CUDA, but the code is not tied to NVIDIA hardware:
it builds equally against AMD's HIP and can also run on CPUs
through HIP-CPU.
The relevant components in the repository are the
\texttt{LineRt} orchestrator interface the Lambda-iteration
loop, the transition-information and equilibrium modules,
and the Kratos-side kernels for photon propagation, imaging,
radiative transfer integration, and block data
management. The pipeline is installable as a Python package,
and the Kratos kernels are compiled within the Kratos
build. Reproducibility commands for the validation suite are
provided with the distribution.\footnote{
  The whole \kratos{} toolset, including the pipeline and
  the backend, is publicly available at
  \url{https://github.com/wll745881210/kratos_linerad}.}

\section{Verification and performance}
\label{sec:verification}

A suite of tests validating the proper implementation of the
line radiative transfer and the imaging scheme is presented
in this section, followed by a performance
characterization. Unless otherwise stated, the quoted
agreement levels are dominated by systematic effects; the
Monte Carlo statistical uncertainties, estimated from the
photon count as $\sigma\sim N^{-1/2}$, are smaller than the
quoted tolerances. The mean-depth convention is adopted, in
which the optical depth is
$\tau_m = n_l \sigma_0\pi^{1/2} L_{\rm slab}/2$ for a slab
of half-thickness $L_{\rm slab}$; the reader is referred to
\citet{1990ApJ...350..120N} for the relation between this
and the line-center convention.

\subsection{Escaped spectra: the Neufeld scaling}
\label{sec:verif-neufeld}

\begin{figure}
  \centering
  \includegraphics[width=\columnwidth]{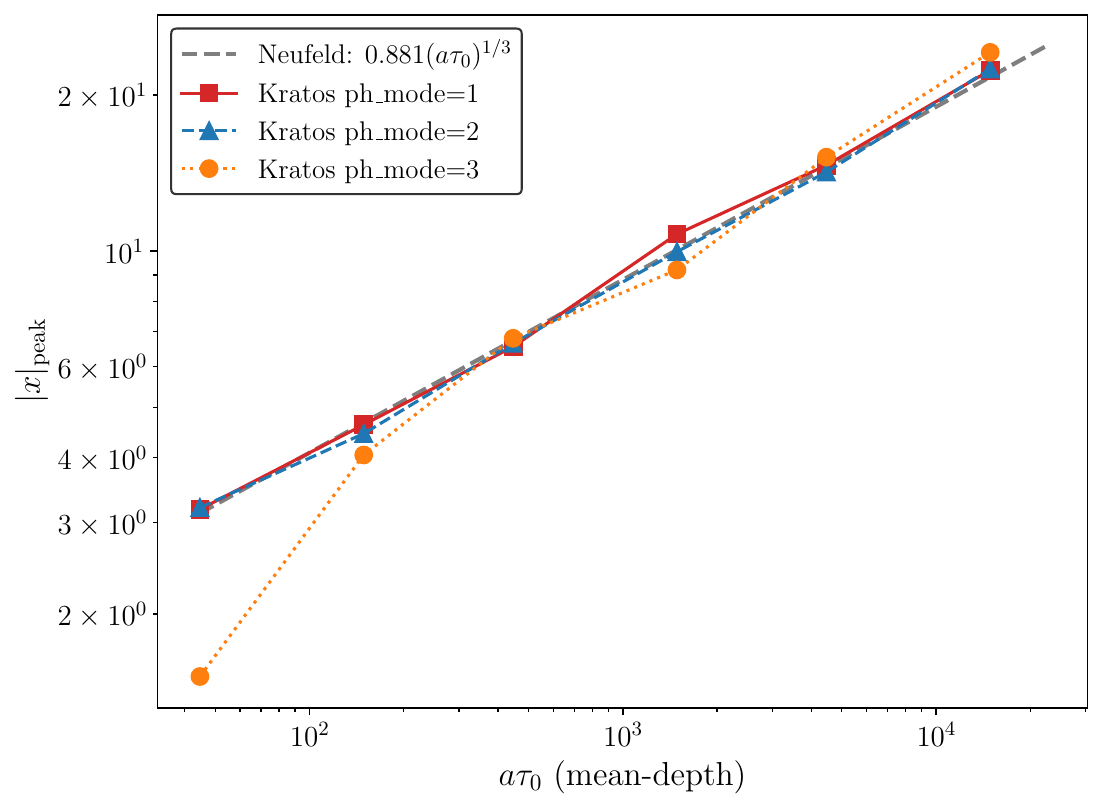}
  \caption{Peak of the escaped frequency distribution versus
    the effective mean-depth optical depth $a\tau_0$. See
    \S\ref{sec:method-mc} for the photon packet modes, and
    the dashed line is the analytic scaling
    $0.881 (a\tau_0)^{1/3}$ \citep{1990ApJ...350..120N}.
  }
  \label{fig:neufeld-xpeak}
\end{figure}

\begin{figure*}[!t]
  \centering
  \includegraphics[width=\textwidth]{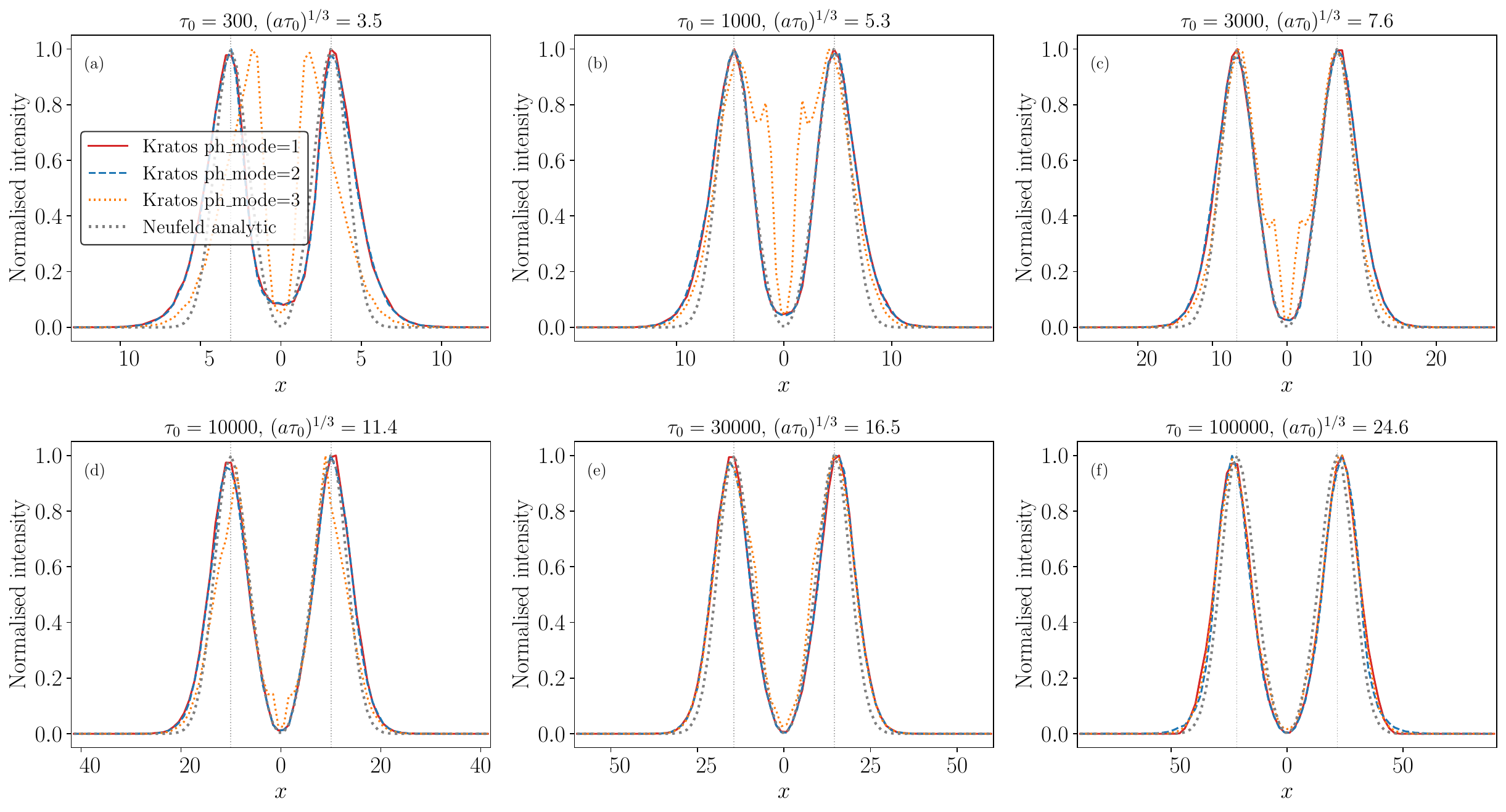}
  \caption{Emergent spectra of the \citet{1990ApJ...350..120N}
    test at each mean optical depth $\tau_0$, normalized to
    the peak.  The dotted gray curves are the analytic
    emergent intensity of \citet{1990ApJ...350..120N}; the
    colored curves are the measured emergent spectra with the
    exact redistribution in the global-memory (solid) and
    constant-memory (dashed) variants and with the approximate
    treatment (dotted).  The vertical light gray dotted lines
    mark the analytic peak positions
    $\pm0.881(a\tau_0)^{1/3}$.  The double-peaked profiles and
    the outward shift of the peaks with increasing optical
    depth are reproduced by the exact treatments.}
  \label{fig:neufeld-spectra}
\end{figure*}

The escaped spectrum is first validated against the
analytical scaling of \citet{1990ApJ...350..120N}. A
plane-parallel slab with $128\times2\times2$ cells, periodic
boundaries transverse to the slab normal, a central
isotropic source of line-center photons, and a damping
parameter $a = 0.149$, characteristic of a strongly damped
resonance line, spanning mean optical depths
$\tau_0 \in \{200, 500, 2000, 8000, 32000\}$ is propagated
with $10^{5}$ photon packets. The test exercises all three
redistribution implementations: the exact kernel table in
global memory, the constant-memory variant, and the
approximate treatment. The core statistics measured are the
the peak locations of the double-peaked profile, whose
analytic scaling is
$x_{\rm peak} = 0.881\, (a \tau_0)^{1/3}$.
Figure~\ref{fig:neufeld-xpeak} compares the measured escape
peak against this scaling, and
Figure~\ref{fig:neufeld-spectra} shows the emergent spectra.
The exact treatments agree with the analytic scaling to
within a few percent across the full range of optical
depths, and the two memory variants are consistent at the
one to $\sim 2\%$ level.

\subsection{Scattering with absorption}
\label{sec:verif-abs}

The two-step scheme decouples absorption from scattering:
the Monte Carlo pass samples the scattering emissivity
without absorption, and the imaging pass applies the
absorption in the formal solution. This separation is
validated against the dusty-slab test of
\citet{2006A&A...460..397V}, in which a uniform static slab
at $T=10$~K ($a=0.015$) with line-center optical depth
$\tau_0=10^{5}$ and a central plane of monochromatic
line-center photons is embedded with dust at seven
absorption depths. The escape fraction $f_{\rm esc}$ is
measured from $2\times10^{5}$ photon packets per point.

\begin{figure}
  \centering
  \includegraphics[width=\columnwidth]
  {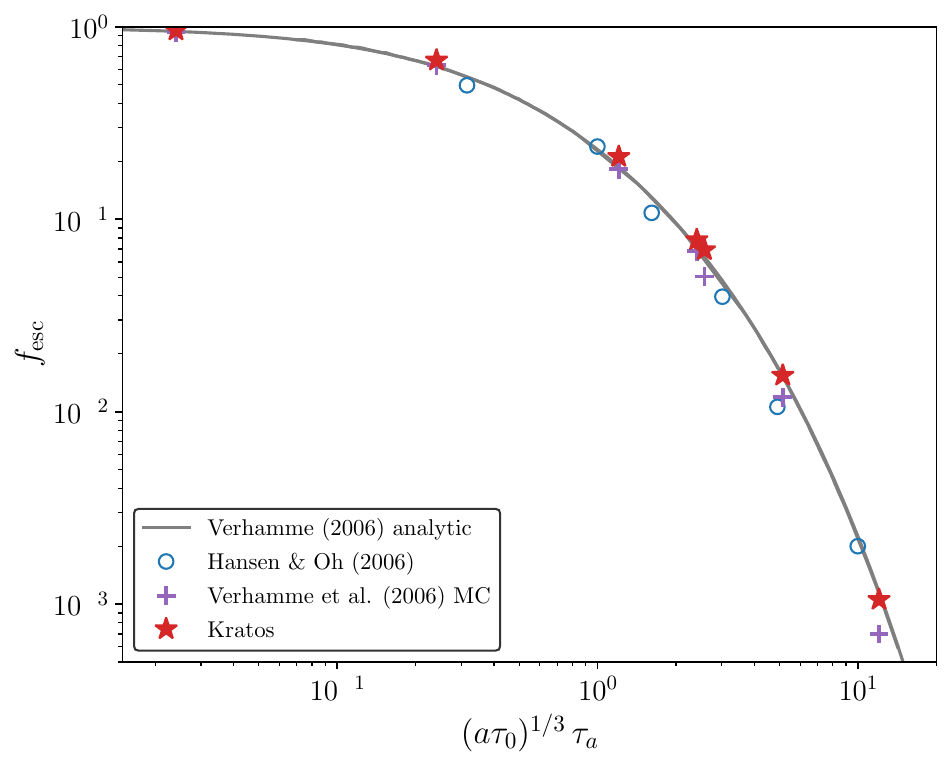}
  \caption{Escape fraction $f_{\rm esc}$ versus the
    dimensionless absorption parameter
    $(a\tau_0)^{1/3}\tau_a$ for the dusty-slab test of
    \citet{2006A&A...460..397V}: a uniform static slab at
    $T=10$~K ($a=0.015$) with line-center optical depth
    $\tau_0=10^{5}$, with monochromatic line-center photons
    emitted from the central plane. The gray curve is the
    analytic reference of \citet{2006A&A...460..397V}, the
    purple crosses are their published Monte Carlo results,
    the open blue circles are the numerical results of
    \citet{2006MNRAS.367..979H}, and the red stars are the
    \kratos{} measurements. 
  }
  \label{fig:absscat}
\end{figure}

Figure~\ref{fig:absscat} shows $f_{\rm esc}$ as a function
of the dimensionless parameter $(a\tau_0)^{1/3}\tau_a$,
which measures the dust absorption depth relative to the
frequency-diffusion length scale: a photon escapes the line
only after random-walking in frequency to a point where the
wing opacity has fallen below the dust opacity, and the
characteristic escape frequency scales as
$(a\tau_0)^{1/3}$. The \kratos{} measurements track the
analytic reference curve of \citet{2006A&A...460..397V} to
within $\simeq12$~per cent across nearly three decades in
$f_{\rm esc}$, from the nearly unabsorbed limit
$f_{\rm esc}\simeq1$ down to $f_{\rm esc}\sim10^{-3}$. The
published Monte Carlo crosses of \citet{2006A&A...460..397V}
scatter below the curve at the deepest absorption depths,
where their runs were photon-starved; the \kratos{} points
remain smooth and close to the analytic reference
throughout, and are consistent with the independent
numerical results of \citet{2006MNRAS.367..979H}. This
agreement confirms that the absorption--scattering
decoupling in the two-step scheme is accurate from the
optically thin to the optically thick absorption regime.

\subsection{Imaging of optically thin moving slab}
\label{sec:verif-thin}

The imaging pass is first validated in the optically thin
regime, where the formal solution reduces to a direct volume
integration of the emissivity and the entire chain,
namely redistribution into the camera direction, bulk-velocity
coupling, and the conversion of photon weights into physical
intensities, is checked against a closed-form spectrum. A
uniform scattering slab of thickness $L_{\rm slab}$ and
line-center optical depth $\tau_0 = 0.01$, with a near-Doppler
profile ($a = 0.01$, $b = 1$~km~s$^{-1}$), is placed
in a cubic domain with free boundaries, illuminated by a
plane-parallel beam of line-center photons propagating in
the plane of the slab, and viewed face-on by a camera
looking along the slab normal, so that the scattering angle
between the incoming photon direction and the line of sight
is $90^\circ$ ($g = 0$). The slab moves at $v_z = b$ along
the line of sight.

For perpendicular scattering in the Doppler limit the
redistribution kernel loses all frequency memory: the
outgoing frequency is drawn from the thermal distribution
independently of the incoming one,
$R(x_{\rm out}; x_{\rm in}, 0) = \pi^{-1/2}\exp(-x_{\rm
  out}^2)$, and the emergent spectrum of the
singly-scattered beam is the redistribution function itself,
\begin{equation}
  \label{eq:thin-slab}
  I(v) = \frac{F \tau_0}{4 \pi b \pi^{1/2}}\,
         \exp(- x_{\rm out}^2)\ ,
  \qquad
  x_{\rm out} = \frac{v + v_z}{b}\ ,
\end{equation}
with the frequency-integrated intensity
$\int I\, {\rm d}v = F \tau_0 / (4 \pi)$. The bulk
motion shifts the line to $v = -v_z$ in the channel
convention of the imaging pass.

\begin{figure}
  \centering
  \includegraphics[width=\columnwidth]
  {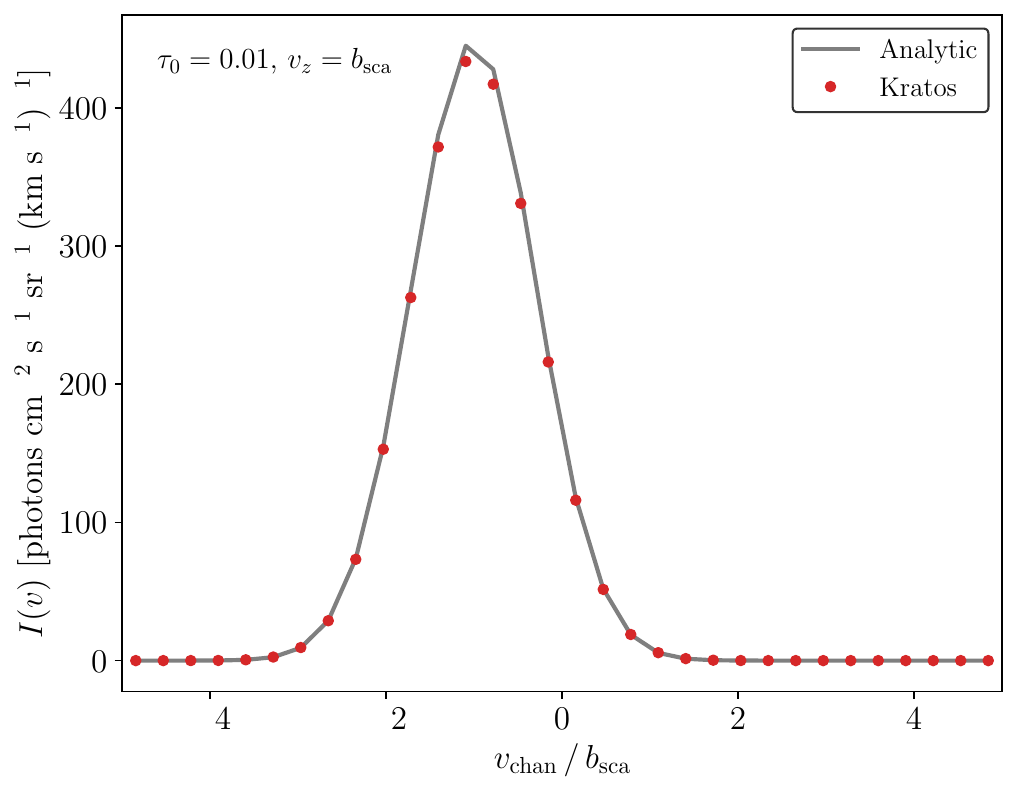}
  \caption{Spectrum of an optically thin moving slab
    ($\tau_0 = 0.01$, near-Doppler profile ($a = 0.01$)
    with $b = 1$~km~s$^{-1}$, bulk velocity $v_z = b$ along
    the line of sight), illuminated by a plane-parallel beam
    in the plane of the slab and viewed face-on.
  }
  \label{fig:thin-slab}
\end{figure}

The measured channel spectrum agrees with the analytic shape
to within $\simeq0.4$~per~cent and the integrated intensity
to within $\simeq2$~per~cent with $2\times10^{5}$ photon
packets (Figure~\ref{fig:thin-slab}), and the line peaks at
$v = -v_z$ as expected; the residual deviations are
dominated by the channel binning of the thermal width ($32$
channels spanning $\pm5\,b$) and the Monte Carlo statistics.
The absolute agreement, without any normalization freedom,
confirms the intensity calibration of the imaging pass.

\subsection{Imaging of optically thick profiles}
\label{sec:verif-imaging}

\begin{figure*}[!t]
  \centering
  \includegraphics[width=\textwidth]
  {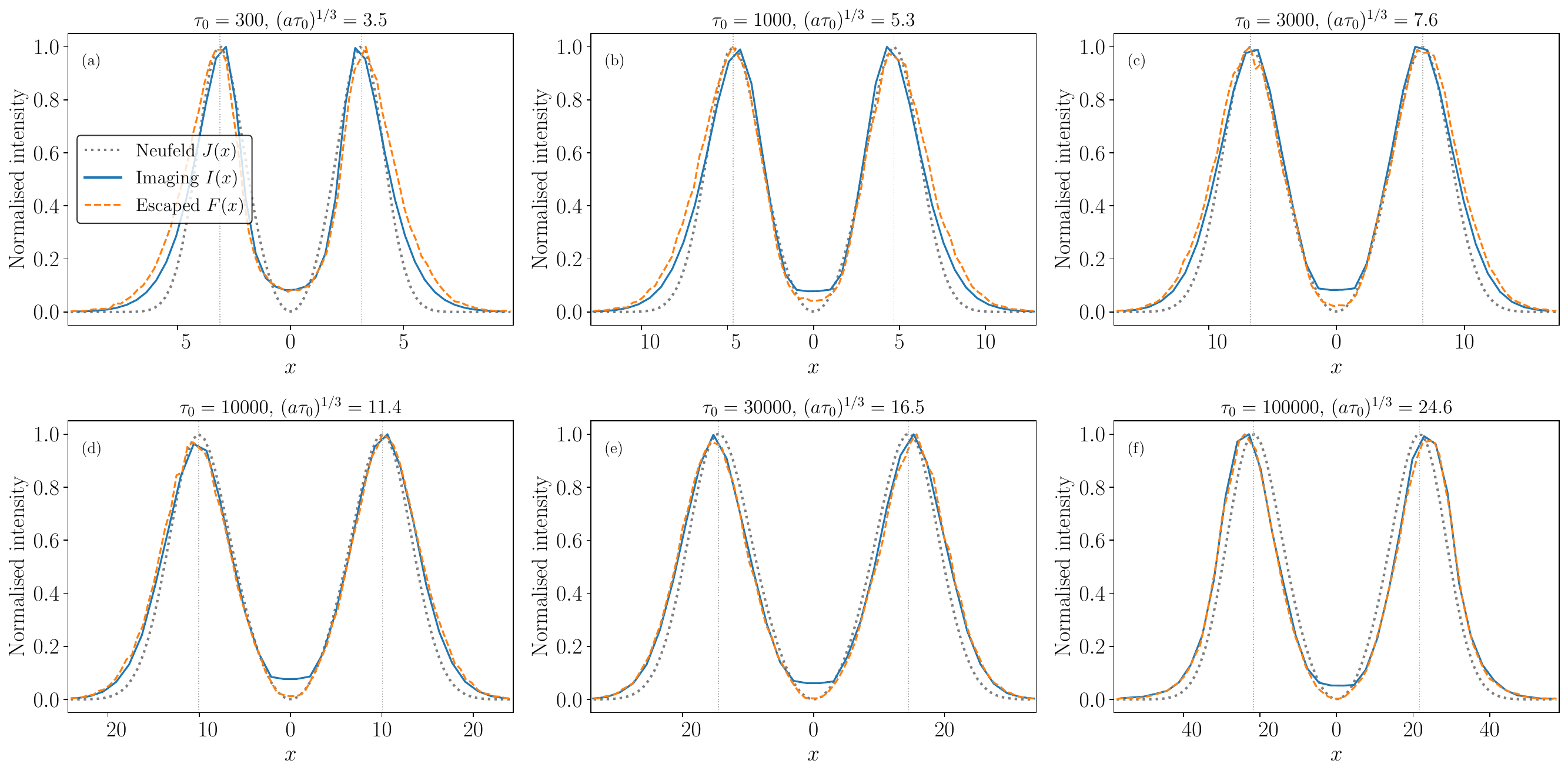}
  \caption{Similar to Figure~\ref{fig:neufeld-spectra}, but
    for the spectra synthesized by the two-step imaging
    scheme (blue curves), shown together with the
    escaped-photon spectra (orange dashed) and the analytic
    emergent intensity (gray dotted).
  }
  \label{fig:imaging-spectra}
\end{figure*}

\begin{figure}
  \centering
  \includegraphics[width=\columnwidth]
  {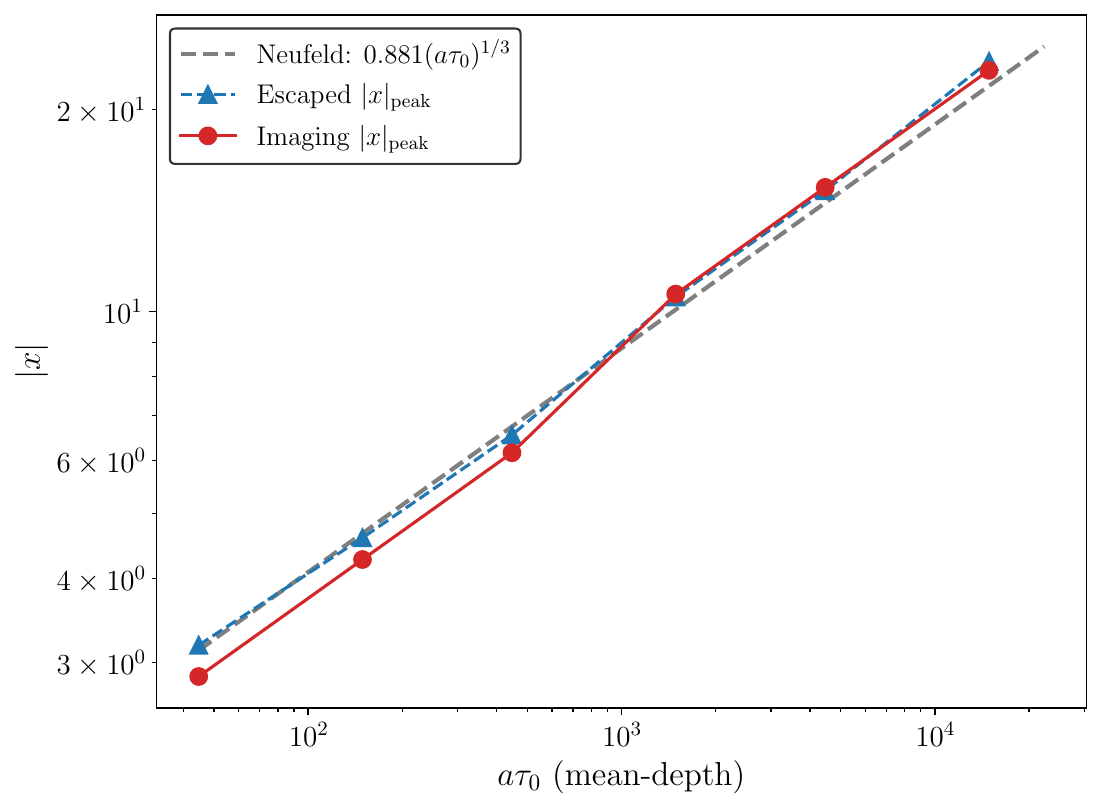}
  \caption{Similar to Figure~\ref{fig:neufeld-xpeak}, but
    presenting the scaling relations from the multi-channel
    imaging results: the imaging peaks (red circles) are
    compared with the peaks of escaped photons (blue
    triangles).  The gray dashed line is the analytic
    scaling $0.881(a\tau_0)^{1/3}$. 
  }
  \label{fig:imaging-peaks}
\end{figure}

The imaging module is also validated by comparing the peak
of the velocity-resolved channel map against the same
Neufeld scaling.  A camera is placed on the $+x$ face with
$\theta=\pi/2$ and $\phi=0$, using $64$ uniformly spaced
velocity channels whose range adapts to the expected escape
width, spanning mean optical depths
$\tau_0 \in \{300, 1000, 3000, 10000, 30000, 100000\}$ with
$10^{5}$ photon packets per point. The imaging peaks
reproduce the Neufeld prediction to within roughly $10\%$
across the tested range of optical depths, with the imaging
peak lying slightly inside the escape peak at the lowest
optical depths and the two converging at the highest. This
offset is physical rather than numerical. The
imaging pass reports the emergent intensity along the camera
direction, and the contribution function $e^{-\tau}$ in the
formal solution weights this intensity toward the layer in
which the optical depth along the line of sight at the given
frequency is of order unity: the imaging spectrum therefore
reflects the source function of that last-scattering layer,
redistributed once into the camera direction. The
escaped-photon spectrum, in contrast, is the
direction-integrated flux of the full escaping ensemble,
which also contains photons that escaped at oblique angles
after longer slant paths and correspondingly more
scatterings; these photons have diffused further in
frequency and displace the escaped peak slightly outward. At
high optical depth, escape is dominated by single excursions
into the far wing, the last scattering is nearly isotropic,
and the two spectra converge.
Figure~\ref{fig:imaging-spectra} compares the imaging
spectra with the analytic profiles, and
Figure~\ref{fig:imaging-peaks} summarizes the peak
positions.

\subsection{Performance tests}
\label{sec:verif-perf}

\subsubsection{Time consumption decomposed}

The wall-clock performance of the two stages is summarized
as follows. Sampling the scattering emissivity with
$10^{5}$ photon packets and up to $10^{7}$ propagation
steps, with exact redistribution and no core-skipping,
completes in roughly $\lesssim 2~\s$ on a single NVIDIA
RTX~3090 GPU. This is about $15\times$ faster than the
eight-core CPU reference (AMD Ryzen~7 5800X) running the
same propagation. The
subsequent ray-tracing imaging stage synthesizes a
$256\times 256$-pixel map with ten velocity channels in
$\lesssim 3~\s$, and this cost is incurred only after the
population iteration has converged. Because the imaging
stage is decoupled from the scattering optical depth and
from the Monte Carlo packet count, its cost does not grow
with the complexity of the scattering problem.


It is emphasized that the performance claim here is a
capability statement rather than a head-to-head ratio
against a specific third-party code. Comparisons should be
decomposed because the relevant quantity is the cost of the
combined iterative population plus imaging loop, and because
existing GPU line codes typically adopt a peeling-off
imaging method with fixed populations rather than the
two-step scheme presented here.  Instead of quoting a single
speedup ratio, the costs are characterized component-wise,
with photon sampling and imaging separated, which is the
comparison that isolates the algorithmic advantage of the
two-step method.

\subsubsection{Comparison of speed with SKIRT}
\label{sec:verif-skirt}

The component-wise costs of \S\ref{sec:verif-perf} are now
complemented by a controlled, decomposed benchmark. Such
comparisons are meaningful only when the physics, geometry,
and source are matched exactly and all problem-specific
accelerations are disabled; this configuration was
constructed against SKIRT~9 \citep{2020A&C....3100381C}, a
widely used CPU Monte Carlo radiative transfer code,
operated in its \lya transfer mode with all
acceleration schemes turned off. The setup is a uniform
cubic medium on a $32^3$ Cartesian grid at $T=100$~K
($b=1.285$~km~s$^{-1}$, damping parameter
$a=4.73\times10^{-3}$), with line-center optical depth
$\tau_0=10^3$ and a uniform volume source emitting a
Gaussian line profile of the thermal width; \kratos{} was
configured identically, with exact $R_{\rm IIA}$
redistribution (\S\ref{sec:method-redistribution}). The
escaped spectra of the two codes are consistent: both show
double peaks at $|x|\simeq2.4$--$2.6$, matching to
$11$~per~cent, and both lie above the plane-parallel
mean-depth prediction $0.881(a\tau_0)^{1/3}=1.48$
\citet{1990ApJ...350..120N}, as expected for a box geometry
in which photons escape in six directions rather than two.
The residual $11$~per~cent difference in the peak position is
systematic rather than statistical, the Monte Carlo
uncertainties being much smaller; it most plausibly arises
from the different spatial discretizations and the
interpolation of the scattering emissivity onto the respective
grids, which slightly shift the effective last-scattering
surface.

Table~\ref{tab:skirt} gives the wall-clock scaling with the
photon-packet count $N_{\rm ph}$. Two times are reported for
each code: the Monte Carlo transport kernel time, and the
end-to-end wall time including fixed overheads. The
\kratos{} overhead is a fixed $\sim0.3$~s per run,
dominated by the one-time GPU construction of the $R_{\rm
IIA}$ redistribution kernel table (which cost $\sim3$~s
when assembled on the CPU), whereas the SKIRT startup cost
is below $0.1$~s. The transport kernel is $6\times$ faster
at $N_{\rm ph}=10^3$ and $132\times$
faster at $N_{\rm ph}=10^6$, saturating at $155\times$ at
$N_{\rm ph}=10^7$, reflecting the
throughput contrast between a single RTX~3090 GPU and a
16-thread CPU. The end-to-end ratio climbs from $0.5\times$
at $N_{\rm ph}=10^3$ to $112\times$ at $N_{\rm ph}=10^7$,
the fixed overhead being significant only at the lowest
photon counts, where the GPU kernel-launch cost dominates
the \kratos{} wall time. The benchmark measures the
Monte Carlo transport stage alone; the deterministic imaging
stage (\S\ref{sec:method-imaging}) is not part of the
comparison, as the SKIRT instrument model differs from the
ray-tracing scheme used here.

\begin{deluxetable*}{rcccccc}
  \tabletypesize{\scriptsize}
  \tablecaption{Performance benchmark of \kratos{} against
    SKIRT~9}
  \label{tab:skirt}
  \tablehead{\colhead{} & \multicolumn{2}{c}{Kratos${}^*$} &
    \colhead{} & \multicolumn{2}{c}{SKIRT${}^\dagger$} &
    \colhead{}
    \\
    \cline{2-3} \cline{5-6} \colhead{$N_{\rm ph}$} &
    \colhead{MCRT${}^{**}$ (s)} &
    \colhead{Total${}^{\ddagger}$ (s)} & &
    \colhead{MCRT${}^{**}$ (s)} &
    \colhead{Total${}^{\ddagger}$ (s)} & \colhead{MCRT
      speedup}} \startdata
  $10^3$ & $0.030$ & $0.42$ & & $0.19$ & $0.2$ & $6$ \\
  $10^4$ & $0.034$ & $0.31$ & & $0.60$ & $0.6$ & $18$ \\
  $10^5$ & $0.086$ & $0.38$ & & $5.90$ & $5.9$ & $69$ \\
  $10^6$ & $0.435$ & $0.84$ & &  $57.60$ & $57.6$ & $132$ \\
  $10^7$ & $3.901$ & $5.44$ & &  $606.50$ & $606.6$ & $155$ \\
  \enddata \tablecomments{ Monte Carlo simulations conducted
    with identical setups (\lya photons, uniform box,
    $\tau_0=10^3$, $32^3$ grid, volume source, exact
    redistribution, no acceleration), for wall-clock times
    versus photon-packet counts. \\
    $*$: On a single NVIDIA RTX~3090 GPU.\\
    $\dagger$: On an AMD Ryzen~7 5800X CPU with 8 cores and 16 threads. \\
    $**$: Monte Carlo transport kernel time (including GPU
    kernel launching overheads).\\
    $\ddagger$: Including fixed overheads ($\sim0.3$~s for
    \kratos{}, dominated by the one-time $R_{\rm IIA}$
    kernel-table construction, now assembled on the GPU and
    amortized over multiple runs; $\lesssim0.1$~s startup for SKIRT).}
\end{deluxetable*}

\subsection{Application example: on a molecular cloud
  simulation}
\label{sec:demo-ism}

As an application example, the pipeline is applied to a
snapshot of the fiducial turbulent diffuse molecular cloud
simulation of \citet{2024ApJ...973...37Y}. The simulation
follows three-dimensional, non-ideal magnetohydrodynamics
coevolved with nonequilibrium thermochemistry (a 23-species
network) with \textsc{Athena++} in a periodic
$(0.04~{\rm pc})^3$ box on a uniform $128^3$ grid, including
the interstellar radiation field at $0.3\,G_0$ and
cosmic-ray ionization. The snapshot analyzed here has a mean
gas temperature $\langle T\rangle\simeq47$~K (peak
$\simeq610$~K), mean densities
$\langle n_{\rm CO}\rangle\simeq2.8\times10^{-5}$~cm$^{-3}$
and
$\langle n_{\rm OH}\rangle\simeq5.8\times10^{-6}$~cm$^{-3}$,
and a line-of-sight velocity dispersion
$\sigma_v\simeq0.5$~km~s$^{-1}$.

Two transitions are synthesized: the OH 18~cm ground-state
$\Lambda$-doublet line at 1665.402~MHz (the 18~cm ground state
is a hyperfine quartet at 1612, 1665, 1667, and 1720~MHz; we
treat the 1665~MHz main line as a single transition,
ignoring the hyperfine splitting and the conjugate behavior
of the main-line pair), the only molecular
line in such diffuse gas that approaches unit optical depth,
with a mean line-center optical depth
$\langle\tau\rangle\simeq0.3$ rising to $\tau\simeq1$ along
the densest sightlines, and the CO $J=1\to0$ line at
115.27~GHz, which remains optically thin
($\tau\simeq0.05$). Both are modeled as two-level systems
with LTE level populations at the local gas temperature,
exact $R_{\rm IIA}$ redistribution with the constant-memory
kernel table (\S\ref{sec:method-gpu-constmem}), and no
continuum opacity.  Channel maps are generated with the
two-step imaging scheme (\S\ref{sec:method-imaging})
on $128^2$ pixels with five velocity channels of width
$\Delta v=1$~km~s$^{-1}$ spanning $v\in[-2,2]$~km~s$^{-1}$,
viewed along the $x$-axis of the simulation box (a single
viewing angle); the Monte Carlo pass used
$\simeq2.1\times10^6$ photon packets per transition.

\begin{figure*}[!t]
\centering
\includegraphics[width=\textwidth]{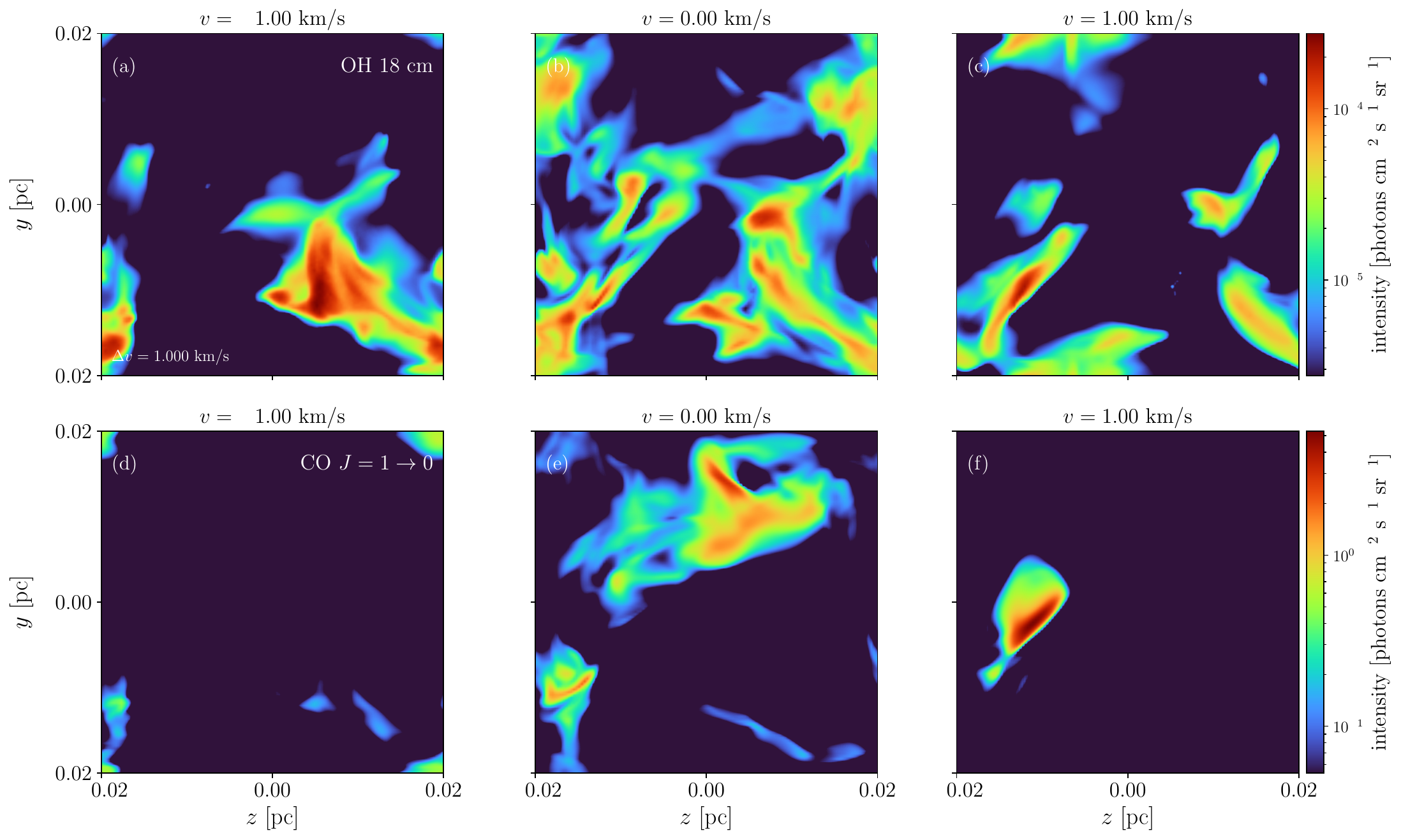}
\caption{Channel maps of the turbulent diffuse molecular
  cloud simulation of \citealt{2024ApJ...973...37Y},
  synthesized with the two-step imaging scheme
  (\S\ref{sec:method-imaging}).  Top row: the OH 18~cm
  $\Lambda$-doublet line at 1665.402~MHz.  Bottom row: CO
  $J=1\to0$ at 115.27~GHz. Columns show the $v=-1$, $0$, and
  $+1$~km~s$^{-1}$ channels of a five-channel deck with
  $\Delta v=1$~km~s$^{-1}$; each row shares a common
  logarithmic intensity scale spanning two decades below the
  row maximum. The maps have $128^2$ pixels over the
  $(0.04~{\rm pc})^2$ face of the simulation box, viewed
  along the $x$-axis.}
\label{fig:ism-channels}
\end{figure*}

Figure~\ref{fig:ism-channels} shows the central three
channels; the outermost channels ($v=\pm2$~km~s$^{-1}$)
carry $\lesssim1$~per cent of the flux and are omitted. The
two lines trace morphologically distinct structures,
reflecting the different regions relatively abundant in CO
and OH in these simulations \citep{2024ApJ...973...37Y}.
At relatively low optical depths, the velocity-integrated CO
and OH intensities correlate with the true column densities
with correlation coefficients $r\simeq0.90$ and $0.97$ in 
logarithmic space, respectively.
The channel maps also recover the cloud kinematics, in which
the emission centroid sweeps by $\sim0.01$~pc across the box
between the $-1$ and $+1$~km~s$^{-1}$ channels.
The cost of obtaining these channel maps remains modest:
each deck of images, one Monte Carlo pass followed by the
deterministic ray-tracing pass through all channels, took
$\simeq 12~\s$ of wall-clock time per transition on a single
NVIDIA RTX~3090 GPU including the IO overheads.

\section{Discussion}
\label{sec:discussion}


We place \kratos{} in the context of existing codes and
discuss the design choices and the current limitations of
the method. Specifically, the two-step imaging scheme used
here for lines was introduced for polarized continuum
radiative transfer by \citet{2025arXiv251201283Y}. The two
codes address different physics.  The continuum code treats
polarization by aligned dust grains, whereas \kratos{}
treats frequency-redistributed line transfer.
The methodological connection is the shared two-step
principle, and the line case requires a genuinely new
treatment because the scattering emissivity is frequency-resolved
and velocity-resolved rather than a scalar per cell.

\subsection{Core-skipping: adopting or not}
\label{sec:disc-coreskip}

Many resonant line codes accelerate propagation at high
optical depth by skipping photons across the optically thick
core of the profile, using a criterion such as
$x_{\rm crit} = 0.2\, (a\tau_0)^{1/3}$. \kratos{}
deliberately omits this optimization. The reason is that the
core-skipping criterion requires an accurate estimate of the
local $a\tau_0$, which in a realistic, clumpy medium with a
spatially varying velocity field is not reliably known a
priori; an incorrect estimate either wastes propagation in
the core or, worse, incorrectly skips photons that would
have scattered. The no-skip choice is therefore a robustness
property rather than an oversight, and it degrades
gracefully, being inefficient only at extreme optical depths
where the computational cost is dominated by the core
crossings. For such regimes, the implementation of discrete
diffusion Monte Carlo (DDMC) is planned, in which cells are
switched from transport to diffusion when their optical
depth exceeds a threshold, a technique with a long history
in neutral-particle and photon transport. DDMC is
complementary to the two-step imaging scheme and would
extend the applicability of \kratos{} to the highest optical
depths without compromising the robustness that motivates
the current design.

\subsection{Limitations and future work}
\label{sec:disc-limitations}

The principal current limitation is the split between the
GPU-resident Monte Carlo transport and the Python-side
orchestration. The population iteration, the photon
generation, and the file IO are handled by the Python
pipeline, and their combined overhead can exceed the GPU
transport kernel time by one to two orders of magnitude: the
fixed pipeline cost of $\sim9$~s per run in
\S\ref{sec:verif-skirt}, dominated by file IO and process
launch, is the visible part of this overhead, and the
CPU-side population solve adds a comparable cost for
multi-level systems. For the benchmark cases presented here
the Monte Carlo propagation still dominates the productive
work, so the current arrangement is acceptable, but porting
these components, the population solver, the photon
pre-processing, and the IO layer, to the GPU is the
highest-impact optimization and is planned.  The absence of
core-skipping makes extremely optically thick media, such as
resonant lines with $\tau_0 \gtrsim 10^{6}$, relatively
expensive, while the planned discrete-diffusion Monte Carlo
extension addresses this regime. Additionally, in more
difficult but realistic situations, mixing and blending of
lines, the current implementation targets single-species
line transfer, and the treatment of multiple interlacing
transitions is a natural extension.


We also note that the two-step architecture is well
suited to forward-model fitting. Once the scattering emissivity
has been sampled, re-synthesizing images for new channel
definitions or spatial binning is computationally cheap,
while a new viewing geometry requires re-accumulating the
camera-directed scattering emissivity in a new Monte Carlo
pass. Wiring \kratos{} into a Markov chain Monte Carlo
driver, in which a full Monte Carlo re-sampling is triggered
when the trial parameters alter the scattering emissivity, would
support efficient fitting procedures against observed
channel maps at a fixed viewing geometry. The velocity-field
treatment is not considered a limitation, since the
velocity-space sampling already accounts for bulk motions.

\section{Summary}
\label{sec:disc-summary}

In this paper we have presented \kratos{}, a GPU-accelerated
Monte Carlo line radiative transfer code. Its principal
contribution is the extension of the two-step
scattering-emissivity imaging scheme to line
transfer. A Monte Carlo pass samples a velocity-resolved
scattering emissivity through the $R_{\rm IIA}$
redistribution kernel, and a deterministic ray-tracing pass
synthesizes channel maps decoupled from the scattering
geometry. This decoupling makes iterative statistical
equilibrium and high-resolution velocity imaging jointly
tractable, a combination that is not available in existing
direct-counting GPU codes with fixed populations and that
remains computationally demanding for self-consistent CPU
codes. The code is validated against the
analytical scaling of \citet{1990ApJ...350..120N},
reproduces the imaging double-peak profiles, and recovers
the thin-slab normalization and the escape fractions to high
precision. By leveraging GPU parallelism, mixed precision,
and constant-memory sampling tables, \kratos{} samples the
scattering emissivity at $10^{5}$ packets in roughly $2$~s on a
single RTX~3090, with imaging synthesized in a few
seconds. Future work includes the porting of the multi-level
equilibrium to the GPU, and the implementation of discrete
diffusion Monte Carlo for the highest optical depths.

\begin{acknowledgments}
  This work made use of the LAMDA database
  \citep{2005A&A...432..369S} and the molecular line lists
  of \citet{2015ApJS..216...15L}. This work is also
  supported by the National Natural Science Foundation of
  China (NSFC) [12573067].
\end{acknowledgments}

\bibliographystyle{aasjournalv7}
\bibliography{refs}

\end{CJK*}
\end{document}